\documentclass[usegraphicx,usenatbib]{mn2e}

\usepackage{verbatim}
\usepackage{color}
\usepackage[normalem]{ulem} 
\usepackage{amsmath} 
\usepackage{times}

\newcommand{\devauc}{De Vaucouleurs'}

\def\eps@scaling{1.0}%

\newcommand{\sn}{$S/N$}
\newcommand{\Msn}{$(S/N)_{\textrm{matched}}$}

\newcommand{\vecg}{\mbox{\boldmath $g$}}
\newcommand{\vecD}{\mbox{\boldmath $D$}}
\newcommand{\vecQ}{\mbox{\boldmath $Q$}}
\newcommand{\matR}{\mbox{$\bf R$}}
\newcommand{\matC}{\mbox{$\bf C$}}
\newcommand{\bnabg}{ \boldsymbol{\nabla_g}}

\newcommand{\desreq}{$4\times 10^{-3}$}
\newcommand{\lsstreq}{$2\times 10^{-3}$}

\newcommand{\sersic}{S\'{e}rsic}

\newcommand{\lognormscatt}{30}

\title{An Implementation of Bayesian Lensing Shear Measurement}

\author[Erin S. Sheldon]{Erin S. Sheldon\thanks{E-mail: erin.sheldon@gmail.com}\\
Brookhaven National Laboratory, Bldg 510, Upton, New York 11973}

\begin{document}

\maketitle

\begin{abstract}

The Bayesian gravitational shear estimation algorithm developed by \cite{ba14}
can potentially be used to overcome multiplicative noise bias and recover shear
using very low signal-to-noise ratio (\sn) galaxy images.  In that work the
authors confirmed the method is nearly unbiased in a simplified demonstration,
but no test was performed on images with realistic pixel noise.  Here I present
a full implementation for fitting models to galaxy images, including the
effects of a point spread function (PSF) and pixelization.  I tested the
implementation using simulated galaxy images modeled as \sersic\ profiles with
$n=1$ (exponential) and $n=4$ (\devauc), convolved with a PSF and a flat pixel
response function.  I used a round Gaussian model for the PSF to avoid
potential PSF-fitting errors. I simulated galaxies with mean {\em observed,
post-PSF} full-width at half maximum equal to approximately 1.2 times that of
the PSF, with log-normal scatter.  I also drew fluxes from a log-normal
distribution. I produced independent simulations, each with pixel noise tuned
to produce different mean \sn\ ranging from $10-1000$.  I applied a constant
shear to all images.  I fit the simulated images to a model with the true
\sersic\ index to avoid modeling biases.  I recovered the input shear with
fractional error $\Delta g/g < 2 \times 10^{-3}$ in all cases.  In these
controlled conditions, and in the absence of other multiplicative errors, this
implementation is sufficiently unbiased for current surveys and approaches the
requirements for planned surveys.

\end{abstract}


\section{Introduction} \label{sec:intro}

Noise bias is one of the largest potential systematic errors in weak
gravitational shear measurements.  For some point estimates of the shear,
such as the maximum-likelihood or expectation value of a galaxy shape, noise can
result in a multiplicative calibration error as well as a PSF-correlated
additive error \citep{HirataAlign04,Refreg12,Melchior12,Miller13}.  Averaging
these point estimators gives a biased estimate of the shear, and the bias
remains even when response terms are included, such as that introduced by
\cite{ksb95} and \cite{Bern02}.

For galaxy images with signal-to-noise ratio (\sn) $\sim 10$, the
multiplicative noise bias can be of order 10\% for galaxies with size
comparable to that of the point-spread-function (PSF).  This multiplicative
error is significantly larger than the requirements for current and planned
surveys, which are $\sim$0.4\% and $\sim$0.2\%, respectively
\citep{HutererSystematics06}.

A number of approaches have been proposed to address noise bias. I do not wish
to give an exhaustive list, but rather a few examples.

A Bayesian approach to shear estimate was proposed by \cite{Miller07}.  With
that method, a point estimator for the shear is still used (the expectation
value of the galaxy ellipticity), but a response term is calculated based on
prior information of the true distribution of galaxy ellipticities and the
likelihood surface for a given galaxy.  A limitation of this method is that no
expression was derived for the mean shear from an ensemble of galaxies, rather
it was proposed to average the responses from individual galaxies.  This method
performs well in comparison to many previous methods, but does not meet the
requirements for current and planned surveys \citep{ba14}.

\cite{Zuntz13} propose to use a point estimate from galaxy shapes and simply
calibrate the answer using simulated data.  A related idea has been proposed in
schematic form by \cite{Refregier13}, to use an iterative approach where the
universe is simulated and its parameters tuned, along with calibration
parameters of the measurement, to reproduce the observed results.  These
methods are in principle limited only by the accuracy of the simulations to
represent the real universe.

A rigorous Bayesian formalism was developed by \cite{ba14}, hereafter BA14,
that has the potential to recover shear with good accuracy using even very low
\sn\ galaxy images.  Rather than relying on a point estimate for the shear,
they derived an expression for the mean shear estimated from an ensemble of
measurements, exploiting the fact that the posterior for the shear must
approach a Gaussian for a large ensemble.  No corrections based on simulations
are required.  They showed that the method is sufficiently unbiased for current
and planned surveys in a simple demonstration with galaxy ellipticities only,
but no implementation for use with galaxy images was presented.

Here I present a full model-fitting implementation of the BA14 algorithm that
works on galaxy images, including the effects of the PSF and pixelization.  I
test the implementation using a set of image simulations with idealized galaxy
models.  To avoid potential PSF-fitting errors, I use only round PSFs, and thus
only test the potential multiplicative bias. I show that in these controlled
conditions the method is sufficiently accurate for current and surveys and
approaches the requirements for planned surveys.

Below I follow the notation of BA14, where the reduced shear was represented by
the glyph \vecg.

\section{Algorithm} \label{sec:algo}

Here I will give a brief overview of the method presented by BA14.  There are
two important assumptions underlying this approach (following the notation in
BA14):

\begin{itemize}

    \item The shear is weak, $\vecg \ll 1$.

    \item The posterior distribution of the mean shear derived from a large
        ensemble of galaxy images is approximately Gaussian.

\end{itemize}

The assumption of small shear is true in many circumstances, but breaks down
along lines-of-sight near large mass over-densities, such as clusters of
galaxies.

The second assumption follows from the central limit theorem: if the shear is
derived from a large enough ensemble of galaxy shapes, the posterior approaches
a Gaussian.  Note this is only true under the assumption that all
galaxies have been sheared equally.

Note no assumptions are made about the likelihood surface for parameters
derived for individual galaxies; only the ensemble average shear from a
population can be assumed to follow a simple Gaussian distribution.

No information is necessarily ``lost'' by restricting shear estimation to
populations rather than individual galaxies.  Galaxies have intrinsic shapes
that are a large effective noise for shear measurement, an order of magnitude
larger than the signal.  Thus some kind of averaging must be performed to
extract a shear from galaxy shapes.  

Given the above assumptions, the authors of BA14 derived a second-order Taylor
expansion of the logarithm of the shear posterior about zero shear, consistent
with the Gaussian assumption:
\begin{eqnarray} \label{eq:pexpand}
    -\ln P(\vecg | \vecD) & \approx & {(\rm const)} - \ln P(\vecg) \nonumber \\
                          & & - \vecg \cdot \sum_i \frac{\vecQ_i}{P_i} \\
    & &
    + \frac{1}{2} \vecg \cdot \left[ \sum_i \left(\frac{\vecQ_i \vecQ^T_i}{P_i^2}
    - \frac{\matR_i}{P_i}\right) \right] \cdot \vecg, \nonumber
\end{eqnarray}
where \vecD\ is the data vector and \vecg\ is the two-component shear.  The
terms $P_i$, $\vecQ_i$, and $\matR_i$ are 
\begin{eqnarray} \label{eq:pqrdef}
P_i     & = & P(\vecD_i | \vecg=0) \nonumber \\
\vecQ_i & = & \left. \bnabg P(\vecD_i | \vecg)\right|_{\vecg=0} \\
\matR_i & = & \left. \bnabg \bnabg P(\vecD_i | \vecg)\right|_{\vecg=0}. \nonumber
\end{eqnarray}
$P$ is the Bayesian ``prior'', and functionally corresponds to the true
distribution of all the relevant galaxy parameters in the absence of shear.
Note the prior on the shear $P(\vecg)$ is assumed to be uninformative.

Accurate prior information for the galaxy shapes is important for this method:
the derivatives of the un-lensed distribution of shapes with respect to shear
encodes how the ensemble of galaxy images responds to a shear. With knowledge of
the true population of shapes, and how that population transforms under shear,
one can infer the applied shear, even in the presence of noise.

To predict the observed distribution of shapes in general, the un-lensed
distribution of shapes must be mathematically sheared and compared to
observables.  However in the approximation given above, only derivatives near
$g=0$ are required and the mean shear $\bar{\vecg}$ and covariance matrix
$\matC_g$ can be found directly:
\begin{eqnarray}
\matC_g^{-1} & = & \sum_i \left(\frac{\vecQ_i \vecQ^T_i}{P_i^2} - \frac{\matR_i}{P_i}\right) \label{eq:cdef} \nonumber \\
\bar{\vecg} & = &  \matC_g \sum_i \frac{\vecQ_i}{P_i}. \label{eq:shdef}
\end{eqnarray}

In practice the parameters of each galaxy are not precisely known. In that case
the derivatives in equation \ref{eq:pqrdef} can be averaged over the full
likelihood surface for each galaxy, taking care to use the prior
distribution given the shear.  These mean values can then be used in the
aggregates shown in Equation \ref{eq:shdef}.

A conceptual outline for measurement of non-constant shear was also given by
BA14, as were third-order tweaks to the second-order perturbations.

A full implementation, fitting to pixelized galaxy images, was not attempted in
BA14.  The basic formalism was shown to work by drawing shapes randomly from an
analytic distribution, adding noise and a small shear, and recovering that
shear using second-order formula.  Because the likelihood surface was not
derived from an image with realistic pixel noise, it may not accurately
represent the challenges of real data.  In the following sections I will show
tests using simulations of galaxy images with pixel noise, which should be a
good first approximation.

\section{Simulation} \label{sec:sim}

In the absence of an absolute calibration source for weak lensing, some
confidence in a shear measurement technique can still be derived using
simulation.  The Third Gravitational Lensing Accuracy Testing challenge
\citep[GREAT3,][]{great3} provides an excellent set of simulations for testing
shear measurement methods.  However, all the provided simulations include
relatively realistic PSF and galaxy properties, features which can introduce
additional errors beyond noise bias.  Going forward, such simulation efforts
will be important, but for this preliminary work I sought more controlled
conditions.

I simulated elliptical \sersic\ \citep{Sersic63} profiles to represent galaxies.
The \sersic\ profile for a round model is given by
\begin{equation}
I(r) \propto \mathrm{exp} \left[ -\left( \frac{r}{r_0} \right)^{1/n} \right].
\end{equation}
For fixed $n$, the full elliptical model has six parameters: two centroid
parameters, two ellipticity parameters, a size parameter and an amplitude, or
flux parameter.  For the size parameter I used the sum of the variance in each
coordinate: $T=<x^2> + <y^2>$.  I used two distinct values of the index: $n=1$
(exponential disks), and $n=4$ (\devauc\ profiles).  

I simulated pairs of galaxy images with identical structural parameters, but
with position angles offset by 90 degrees, in order to cancel the intrinsic
shape noise.  This simulation strategy is generally known as a ``ring
test''\citep{Nakajima2007}.  Using a ring test greatly reduces the number of
simulated images required to reach a desired precision in the recovered shear.
Traditional ring tests use more than two orientations, which can result in
faster convergence to the true shear, but I found the flexibility of working
with pairs to be useful.  

I convolved the models with a round Gaussian PSF with $\sigma = \sqrt{2}$
pixels.  I chose a Gaussian to minimize PSF modeling errors, and made it round
to avoid the potential additive bias associated with a non-circular PSF.  I
then convolved the models with a uniform square pixel response function.

For consistency, I chose the same un-lensed shape distribution
used in \cite{ba14}:
\begin{equation}
P_0(e^s)\propto \left[1-(e^s)^2\right]^2 \exp\left[-(e^s)^2/2\sigma_{\rm prior}^2\right],
\end{equation}
with $\sigma_{\rm prior}=0.3$. This distribution is sufficiently similar to the
distribution of true galaxy shapes to be useful in this test.  This
distribution also has the required property that it is twice differentiable.

I drew the other galaxy parameters from simple non-covariant distributions.  I
drew the size parameter from a log-normal distribution with \lognormscatt\%
scatter.  The mean of the size distribution was chosen such that the average
galaxy image had {\em observed, post-PSF} full-width at half maximum (FWHM)
approximately 1.2 times that of the PSF.  Due to the log-normal scatter, some
galaxies were smaller and some larger.

I ran multiple simulations with different mean signal-to-noise ratios \Msn\ in the
range $[10,1000]$.  Fluxes were drawn from a log-normal distribution and
Gaussian noise was added so that the mean galaxy image had the desired 
\Msn.  The optimal, matched signal-to-noise ratio \Msn\ is defined as
\begin{equation}
    (S/N)^2_{matched} = \frac{1}{\sigma^2_{sky}} \int I^2(x,y) dx dy
\end{equation}
where $I(x,y)$ is the true light profile and $\sigma_{sky}^2$ is the variance
of the Gaussian noise.  This matched filter signal-to-noise can be shown to be
the maximal possible measure of the signal-to-noise ratio, and was used as the
signal-to-noise ratio definition for the GREAT08 and GREAT3 shear measurement
challenges \citep{BridleGREAT08,great3}.  Other measures of \sn\ can be lower
by as much as a factor of two \citep{great3}.

By drawing sizes and fluxes from log-normal distributions I was attempting to
very roughly approximate a selection that might occur in real data, for example
binning galaxies by a robust flux measure such as a PSF flux and performing
star-galaxy separation.  Within a PSF flux bin, the true distribution of flux
would have a small tail to lower values due to noise and a long tail to higher
flux because galaxies can be larger than the PSF, causing the PSF flux to in
some cases be an underestimate.  By tuning the mean FWHM of the galaxies to be
approximately 1.2 times that of the PSF, I was attempting to mimic the sort of
noisy size cut that can occur when separating stars from galaxies at low
signal-to-noise ratio.

Finally, I drew the centroid from a Gaussian in each coordinate, with $\sigma$
of 0.1 pixels.  I was attempting to model the situation in real data where one
may have a poor initial guess at the true centroid that may not agree well with
that derived from the model fit, and the prior on the center would be based on
this guess.  However, this may be sub-dominant to the pixelization effect,
which would result in a more uniform distribution across a
pixel\footnote{Thanks to the anonymous referee for bringing this to my
attention.}.  I did not explicitly account for this distribution in my
simulation.

\section{Implementation} \label{sec:impl}

As described in \S \ref{sec:algo}, in the presence of noise the estimator
involves integrals of $P$, \vecQ\ and \matR\ over the full likelihood surface
for each galaxy.

Exploration of the likelihood surface in a high-dimensional space requires a
large number of model evaluations, so I found efficiency to be important.  As
an optimization, I approximated galaxy models as sums of Gaussians according to
the fits in \citet{HoggGMix}.  I also fit the PSF using a Gaussian.  Using
Gaussians for both galaxy and PSF models facilitated performing fast analytic
convolutions.  This is a fast and simple alternative to convolutions using
Fourier transforms.

By using a fit to the observed, pixelized PSF to convolve the galaxy models, I
accounted approximately for the convolution by the pixels in the PSF model
itself.  This worked well for the parameters I used, but for a poorly sampled
PSF this approach could result in relatively large errors.

I used a fast approximation to the exponential function for evaluating the
Gaussians in the galaxy models.  Even so, evaluation of the exponential
function was the computational bottleneck for the likelihood evaluation.

I applied priors for the non-shape parameters in order to make exploration of
the likelihood surface more efficient.  The prior on the shapes was not applied
during exploration, because the $P$, \vecQ\ and \matR\ are to be averaged over
the likelihood.  This separation was possibly only because the shapes were not
covariant with the other parameters.

During fitting, I used prior distributions matched exactly to the true
distributions used in the simulation.  I used the correct \sersic\ index for
fitting in each case; i.e. when simulating exponential disks, I also fit an
exponential disk ($n=1$).  By choosing the true priors and model family, I
tested the accuracy of the algorithm and likelihood sampling technique directly
and avoided confusion with other issues such as model bias or empirical prior
determination.

I used a Markov Chain Monte Carlo (MCMC) to explore the likelihood surface.  I
found using a standard Metropolis-Hastings algorithm \citep{Metropolis53}
challenging because galaxies have a wide variety of best-fit model parameters
and errors in those parameters, depending on the noise level.  I did not find
it straightforward to predict what the parameter errors would be a priori,
which made it difficult to choose a ``step size'' for the MCMC chain.  This is
one reason I used an affine invariant MCMC \citep{GoodmanWeare10}.  An affine
invariant MCMC is dynamically adapted to the underlying distribution by
comparing multiple ``walkers'' as they are moved through the parameter space;
it thus does not require tuning the step size.  I used the implementation
presented in \citet{Mackey13} for this work.

For the MCMC I used an affine parameter $a=2$, which gives an acceptance rate
of about 0.5. I used 80 ``walkers'' in the chain, with 400 initial steps per
walker for burn-in followed by 200 additional steps per walker for measuring
expectation values.  I found that increasing the number of steps generally
resulted in less bias in the recovered shear, but I found diminishing returns
beyond 200 for the simple models I was fitting.

I chose initial locations for the walkers to be centered on the maximum
likelihood solution found with a Levenberg-Marquardt method \citep{Lev44}. The
location for each walker was chosen from a multi-dimensional Gaussian centered
at the maximum likelihood with scatter based on errors in the fit, truncated to
positive for flux and size and within $(-1,1)$ for shapes.  I also tried
centering the initial positions on the true parameters, as well as drawing the
initial positions from the priors.  All methods gave consistent results.  For
the large number of burn-in steps that I used the result was rather insensitive
to the starting position.

It may be worth while to further optimize the exploration of the likelihood
surface in terms of the number of walkers and steps used, but also the type of
computing resources used.  Graphics processing units may prove particularly
useful; in a preliminary study I found an decrease in time per likelihood
evaluation of order 100 over traditional processors.

As I will show in section \ref{sec:truebias}, the formulas in BA14 break down
at high shear.  However, it is beneficial to perform tests at high shear
because the number of simulated galaxies required to reach a specified
fractional noise in the measured shear goes roughly as the inverse square of
the shear.  For this reason I chose to expand about the true shear instead of
zero shear in Equations \ref{eq:pexpand} and \ref{eq:pqrdef}.  I also performed
a more limited range of tests expanding about zero shear and shearing the
images by $g=0.01$, and got consistent results.   Note, if constant shear is a
valid assumption, the procedure can be iterated, plugging the estimated shear
into the Taylor expansion in each successive iteration.  I found this procedure
to converge in three iterations even for shears of order 10\%. Note this
convergence will not succeed in the presence of large additive biases.

The code is freely available for
download\footnote{https://github.com/esheldon/ngmix}.  Potential users should
note that, at the time of writing, the code is under heavy development and the
application programming interface may evolve rapidly.

\section{Results} \label{sec:results}

\subsection{Calibration Bias vs. True Shear} \label{sec:truebias}

\begin{figure}
 \includegraphics[scale=0.45]{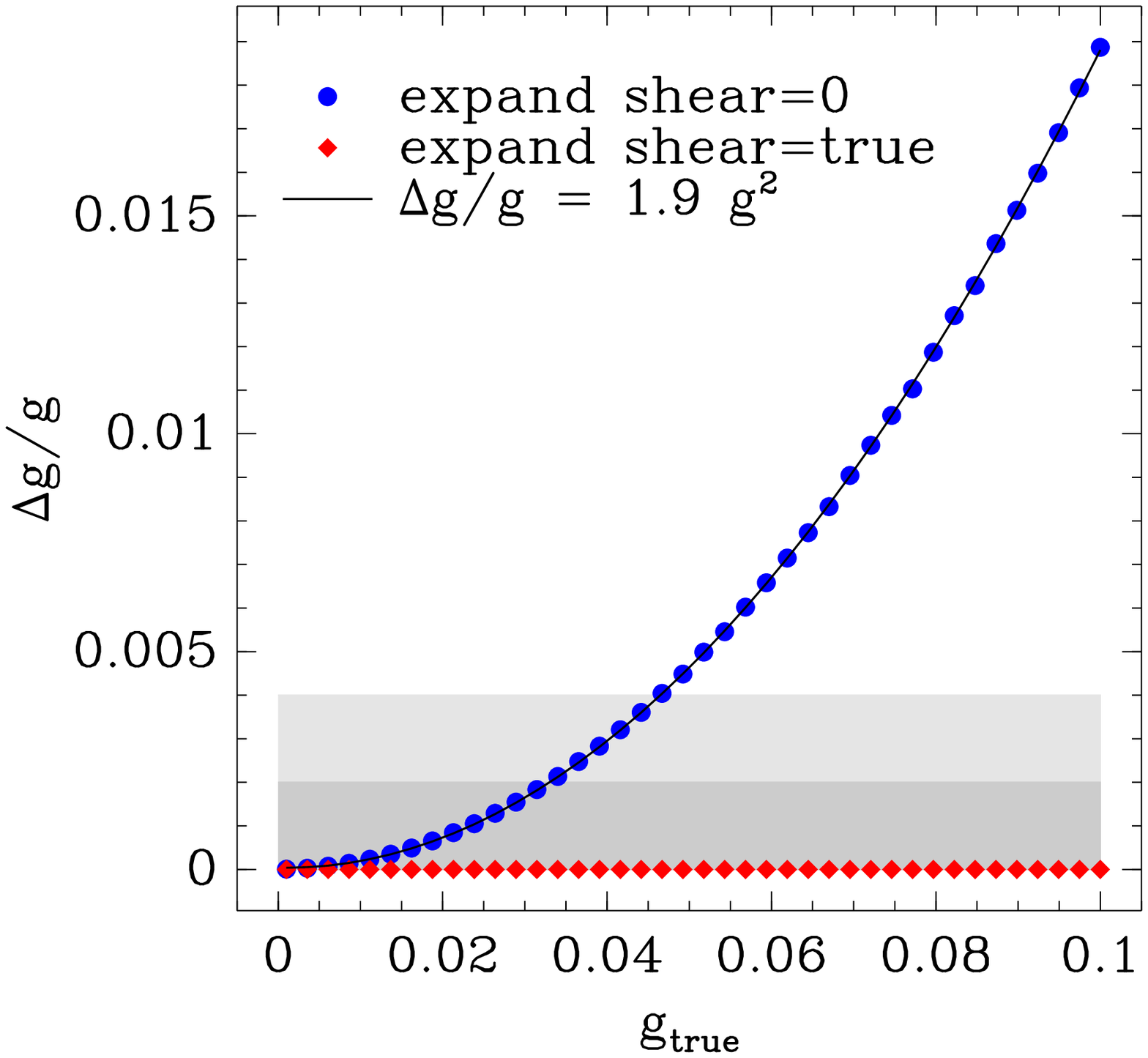}
 \caption{Fractional bias in the recovered shear $\Delta g/g = g/g_{true}-1$
     as a function of true shear,
     in a zero noise simulation.  The blue circles represent the recovered
     shear when using the posterior equations expanded to second order about
     zero shear.  The red diamonds represent expansion about the true shear.
     The solid curve represents the best-fitting quadratic function of the true
     shear $\Delta g/g \sim 1.9~g^2_{true}$.  The quadratic bias as a function of
     true shear indicates a break down of the second-order Taylor approximation
 presented in \citet{ba14}. The light and dark gray bands represent the
 approximate total multiplicative error requirements for current and planned surveys respectively.}
 \label{fig:nonoise}
\end{figure}

Figure \ref{fig:nonoise} contains results for a zero-noise simulation.  The
fractional bias is shown as a function of true shear.  The bias is well fit by
a quadratic function of the true shear, represented by the overlaid curve. This
quadratic error is expected for a second-order Taylor expansion.

Because of this bias, I expanded the equations about the true shear rather than
zero shear, as explained in \S \ref{sec:impl}.

\subsection{Calibration Bias vs. Galaxy Signal-to-noise Ratio} \label{sec:snbias}

\begin{figure}
 \includegraphics[scale=0.45]{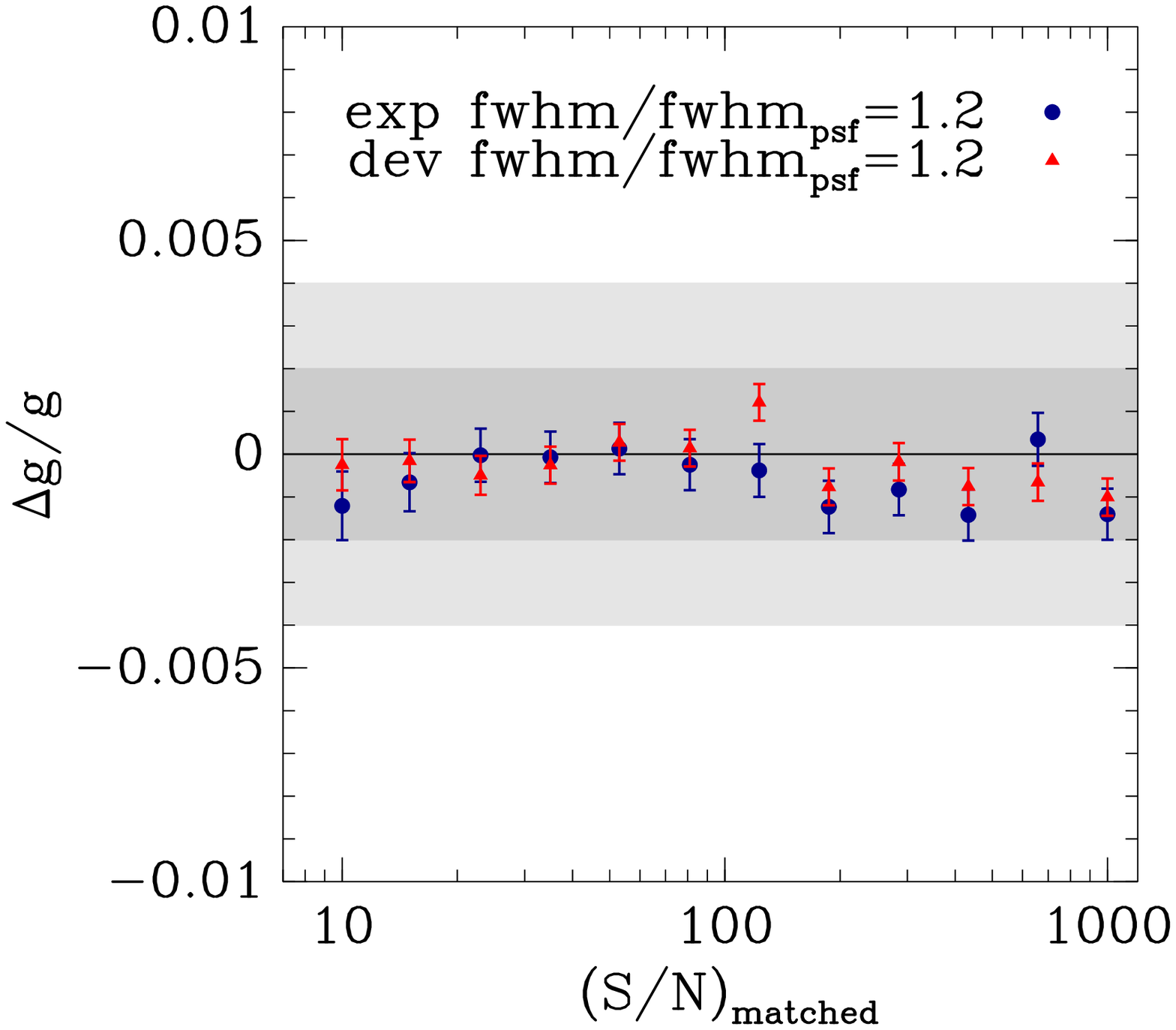}

 \caption{ Fractional bias in the recovered shear for the estimator presented
     in \citet{ba14}.  The bias is plotted as a function of the maximal,
     matched galaxy signal-to-noise ratio \Msn, for galaxies with exponential
     (\sersic\ $n=1$) and \devauc\ profiles ($n=4$).  The simulated galaxy
     parameters were drawn from broad distributions in size, flux, and
     ellipticity.  The size distribution was log-normal with \lognormscatt\%
     scatter and mean such that the {\it observed, post-PSF} FWHM was
     approximately 1.2 times that of the PSF, mimicking small galaxies that may
     pass a size cut, such as that used during star-galaxy separation.  The
     flux distribution was log-normal with \lognormscatt\% scatter, and noise
 chosen to produce the indicated \Msn.  The ellipticities were drawn from the
 simple distribution introduced in \citet{ba14}. The light and dark gray
 bands represent the approximate total multiplicative error requirements for
 current and planned surveys respectively.}

 \label{fig:fracerr}
\end{figure}

Figure \ref{fig:fracerr} contains the fractional error in the recovered shear
as a function of optimal, matched signal-to-noise ratio \Msn, for the simulated
galaxies described in the previous sections.  In these controlled conditions
this implementation is unbiased at the $\Delta g/g \sim 2\times 10^{-3}$ level.

To give a sense of scale, 43,026,754 images were used to estimate the point
at $S/N \approx 10$ for the \devauc\ galaxy test.

The errors for the different values of \Msn\ appear to be correlated.  This may
be due to the details of how the likelihood surface was explored, and may be
correctable by tuning the parameters of the affine-invariant MCMC chain, such
as number of walkers, burn-in, or post-burn-in steps.

\subsection{Comparison with Requirements for Current and Planned Surveys}
\label{sec:req}

In the figures above I showed gray bands representing the approximate
multiplicative bias requirements for current surveys such as the Dark Energy
Survey \citep[][DES]{DESWhitePaper} and the Hyper-Suprime-Cam survey
\citep[][HSC]{HSC12}, and planned surveys such as the Large Synoptic Survey
Telescope survey \citep[][LSST]{IvezicLSST08} and the Euclid Mission
\citep{Euclid2011}.  These requirements are based on the calculations presented
by \citet{HutererSystematics06}. I assumed fiducial sky coverage and that shear
calibration bias results in less than 20\% degradation in the accuracy of the
recovered dark energy equation of state parameter relative to shot noise
errors.  The cut at 20\% is somewhat arbitrary, and was chosen to coincide very
roughly with the stated requirements for DES. The requirements are
$\sim$\desreq\ for current surveys and $\sim$\lsstreq\ for planned surveys.
These requirements are met in all the cases that I tested.  But note that the
quoted requirements represent the {\em total} multiplicative error budget for
these surveys; there may be other sources of bias in addition to noise bias,
such as errors in measuring and interpolating the PSF \citep[e.g.][]{Cropper13},
which may dominate at this level of accuracy.

Smaller and fainter galaxies might have more bias than the average.  The
authors of \cite{ba14} noted that small and faint galaxies, for which the
ellipticity likelihood is broad, get little weight in the average shear
measurement relative to larger and brighter galaxies.  This is an intrinsic
feature of the estimator, no additional weighting is needed.  Indeed I made no
cuts on the simulated galaxies used for this measurement, even though some had
signal-to-noise ratio much less than 10, yet I recovered the shear with good
accuracy.  This is a useful property of the estimator, as the shear measurement
process need not necessarily introduce additional selection effects.

In Figure \ref{fig:nonoise} I showed the multiplicative bias due to the
breakdown of the second-order Taylor expansion at higher shear.   This effect
is prohibitive for current surveys when the shear exceeds 0.05.  The authors of
BA14 suggested potentially using higher order information to tweak the second
order equations.  I have not yet explored that approach.


\section{Discussion} \label{sec:summary}

The success of this implementation in a controlled simulation is strong
encouragement for further development of this method.  This implementation has
significant limitations:  only constant spatially shear can be measured, and at
higher shears some sort of iteration must be used to approach the correct
answer (see \S \ref{sec:impl}).  Furthermore, the model-fitting approach I
presented here may ultimately only be useful as a proof-of-principle:
model-fitting is limited by the accuracy of the models to represent true
galaxies, and thus in real data additional errors will be present; this is the
so-called ``model bias''.  

\cite{Kacprzak13} showed that model bias may be on the order of baseline
requirements for current surveys, but potentially crippling for future surveys
with more stringent requirements. One potential solution is to fit models with
more freedom.  However, fitting a more complex model often results in a more
complex likelihood surface. To fully explore this more complex surface requires
more burn-in for an MCMC chain and more samples to estimate the expectation
values for the $P$, \vecQ\ and \matR\ parameters.  I leave tests with more
complex models to future work.

As an alternative to model fitting, \cite{ba14} propose to measure moments in
Fourier space, making no explicit parameterization of galaxy light
distributions beyond the observed pixel values.  The moments are interpreted
using a Bayesian formalism similar to that used for model fitting, and so could
be robust to noise bias effects.  Evaluation of this Fourier method in the
presence of noise is worthy of future effort.

\section*{Acknowledgments}

ES is supported by DOE grant DE-AC02-98CH10886.

Thanks to Gary Bernstein, Bob Armstrong, Jim Bosch, Mike Jarvis and Robert
Lupton for many useful discussions; thanks to An$\check{\textrm{z}}$e Slosar
for suggesting I expand the equations about the true shear to reduce the number
of simulated images required for each test.  Thanks to the STAR, PHENIX and
LBNE experiments at BNL for use of spare cycles on their compute clusters, and
to the staff of the Rhic Atlas Computing Facility at BNL for their continued,
and disproportionately large support of my work.  Finally, thanks to the
referee for many helpful suggestions.

\bibliographystyle{mn2e}
\bibliography{apj-jour,astroref}

\end{document}